\documentstyle[epsf,floats,prb,aps]{revtex}
\begin{document}
\draft
\title{First-principles and semiempirical calculations for $F$ centers
       in KNbO$_3$}
\author{R.~I.~Eglitis}
\address{
Universit\"at Osnabr\"uck -- Fachbereich Physik,
D-49069 Osnabr\"uck, Germany\\
and Institute of Solid State Physics, University of Latvia,
8 Kengaraga, Riga LV-1063, Latvia}
\author{N.~E.~Christensen}
\address{
Institute of Physics and Astronomy, University of Aarhus,
Aarhus C, DK-8000, Denmark}
\author{E.~A.~Kotomin}
\address{
Institute of Solid State Physics, University of Latvia,
8 Kengaraga, Riga LV-1063, Latvia\\
and Institute of Physics and Astronomy, University of Aarhus,
Aarhus C, DK-8000, Denmark}
\author{A.~V.~Postnikov and G.~Borstel}
\address{
Universit\"at Osnabr\"uck -- Fachbereich Physik,\\
D-49069 Osnabr\"uck, Germany}
\date{Received 19 March 1997}
\twocolumn[\hsize\textwidth\columnwidth\hsize\csname@twocolumnfalse\endcsname
\maketitle

\begin{abstract}
The linear muffin-tin-orbital method combined with density
functional theory (local approximation) and the semiempirical method
of the intermediate neglect of the differential overlap (INDO)
based on the Hartree-Fock formalism are used for the study of the
$F$ centers (O vacancy with two electrons) in cubic
and orthorhombic  ferroelectric KNbO$_3$ crystals. Calculations for 39-atom
supercells show that the two electrons are considerably delocalized
even in the ground state of the defect. Their wave functions extend
over the two Nb atoms closest to the O vacancy and
over other nearby atoms. Thus, the $F$ center in KNbO$_3$ resembles
electron defects in the partially-covalent SiO$_2$ crystal
(the so-called $E^{\prime}_1$ center) rather than usual $F$ centers
in ionic crystals like MgO and alkali halides. This covalency is confirmed
by the analysis of the electronic density distribution.
Absorption energies were calculated by means of the INDO method using
the $\Delta$ self-consistent-field scheme after a relaxation of atoms
surrounding the $F$ center.
For the orthorhombic phase three absorption bands are calculated to lie at
2.72 eV, 3.04 eV, and 3.11 eV. The first one is close to that observed under
electron irradiation. For the cubic phase, stable at high temperatures, above
708 K, only the two bands, at 2.73 eV and 2.97 eV, are expected.
\end{abstract}
]

\section{Introduction}
Due to its technological importance potassium niobate, KNbO$_3$, a
perovskite-type ferroelectric material, has lately been subject to numerous
{\em ab initio} electronic structure calculations. In the cubic KNbO$_3$
crystal each O atom is surrounded by four K atoms, two Nb atoms and eight
next-nearest O atoms. Many of the calculations were based on the
local density approximation (LDA) combined either\cite{ktn3,phonon}
with the linearized muffin-tin orbital (LMTO)\cite{oka} or with
the pseudopotential method \cite{ksv94,loto} as well as with the
linearized augmented plane wave\cite{sb92,chain,s95} (LAPW) scheme.

Complementary to this  approach is the Hartree-Fock (HF) formalism.
Compared to the LDA, the HF scheme has the advantage of the
exact treatment of exchange interactions. Recent implementations
have no restrictions on the spatial form of the potential, no potential
effects due to use of muffin-tin boundary conditions and/or space-packing
empty spheres. It gives the effective charges and suggests a bond-population
analysis between pairs of atoms, and last, it allows one easily to perform
the calculation of excited states and optical absorption energies.

Since  such calculations are quite time consuming, there exist
only a few HF studies for perovskite systems; see, for example, the cluster
calculations of Ref.\onlinecite{donner}. Instead, a simplified
(semiempirical) version of the HF method widely known as intermediate
neglect of the differential overlap\cite{indo1,indo2} (INDO)
has been applied  successfully to calculations for many oxide crystals,
including MgO, \cite{kot2} $\alpha$-Al$_2$O$_3$ (corundum),\cite{kot1}
and TiO$_2$.\cite{sta2} In recent studies
of pure KNbO$_3$ and KTaO$_3$ crystals \cite{egl1,egl2} their electronic
structure and equilibrium ground-state structure for several ferroelectric
phases as well as $\Gamma$-phonon frequencies were reproduced in
surprisingly good agreement with both LDA calculations and available
experimental data.

It is well understood now that {\it point defects} play an important role in
the electro-optic and nonlinear optical applications of KNbO$_3$ and
related materials.\cite{gun1} In particular, reduced KNbO$_3$ crystals
containing oxygen vacancies reveal short-pulse excitations which could be
used for developing fast optical correlators. \cite{zgo} Its use for light
frequency doubling is seriously affected by presence of unidentified defects
responsible for induced IR absorption. \cite{polzik} The photorefractive
effect, important in particular for holographic storage, is also well known
to depend on the presence of impurities and defects.

One of the most common defects in oxide crystals is the so-called $F$ center,
an O vacancy $V_{\mbox{\tiny O}}$  which traps two electrons. \cite{craw} In
electron-irradiated KNbO$_3$ a broad absorption band is observed around 2.7
eV at room temperature and tentatively ascribed to $F$ centers \cite{hod} (see
also Ref. \onlinecite{F+}). This defect is also of theoretical interest for
two reasons.

(i) Due to a low local symmetry of the O sites in the lattice,
the threefold-degenerate $2p$-type excited state could be split  into
several levels responsible for {\it several} absorption bands. This effect
 has been observed a long time ago for the $F^{+}$ centers in corundum,
but theoretically it was examined\cite{sta1} only very recently.
Upon cooling from a high temperature, KNbO$_3$ undergoes a sequence of phase
transitions from a paraelectric cubic phase to ferroelectric tetragonal
phase (at 708 K), then to the orthorhombic structure (at 498 K), and finally
to the rhombohedral (at 263 K) phase. The atomic positions in all these
phases have been determined experimentally. \cite{hewat} Under these phase
transitions the local symmetry of the O vacancy also changes, which can, in
principle, affect the optical properties of the $F$ centers. This problem has
never been addressed.

(ii) Qualitative theoretical analyses of the $F$ centers in perovskites
predict the effect of the symmetry breaking of one-electron orbitals
associated with the {\it asymmetric} electron density delocalization over
the two Nb atoms closest to the O vacancy:
Nb$_1$--$V_{\mbox{\tiny O}}$--Nb$_2$. \cite{pros}

To answer these questions, as well as to check the assignment of the 2.7 eV
absorption band, we study in the present paper the $F$ center in KNbO$_3$
using the supercell model and two different theoretical techniques:
full-potential LMTO and INDO.

The paper is organized as follows. In Sec. \ref{sec:methods} we describe
the essential features of the methods used. In Sec. \ref{sec:res1} we
discuss the main findings obtained for the ground-state properties of the $F$
center while Sec. \ref{sec:opt} describes the absorption energies
calculated by means of the INDO method for two different (the
cubic and orthorhombic) phases. Last, the main results are summarized in
Sec. \ref{sec:conc} where results obtained by the two (quite different)
methods are compared, and the effect of chemical bonding covalency on the $F$
center properties is discussed.

\section{Methods used}
\label{sec:methods}
\subsection{Local density approximation}
Even the crudest approximation, the LDA, to the density functional theory has
been successfully applied to predict structural and dynamical properties of
a large variety of materials. Equilibrium volumes, elastic constants, phonon
frequencies, surface reconstruction, and magnetism are just some examples of
properties which could be successfully calculated for systems without
particularly strong electron correlations within the LDA (or LSDA, the local
spin-density approximation). The LDA usually leads to some overbinding in
solids (equilibrium volumes are typically 1-3 \% underestimated).

The LDA calculations for KNbO$_3$ performed in Ref. \onlinecite{ktn3} yield an
equilibrium volume which is $\approx$ 5\% too small, indicating that the LDA
overbinding in this case is not considerably exceeding the ``acceptable''
limits. This is why in the present paper, although one may expect that
a more accurate treatment of correlation effects may be needed for some
ionic compounds,\cite{fulde95} we apply the LDA to the $F$ center
in KNbO$_3$. The LDA exchange-correlation contribution is accounted for by
means of Perdew and Zunger's parametrization \cite{perzu} of the
calculations by Ceperley and Alder.\cite{ceperl}

The self-consistent solution of the one-electron equation is performed by
means of the LMTO method.\cite{oka} We have used the
``atomic-spheres-approximation''\cite{oka} (ASA) as well as a
``full-potential formalism''\cite{meth} (LMTO-FP). Whereas the LMTO-ASA uses
potentials and charge densities that are made spherically symmetric inside
(slightly) overlapping atomic spheres, no shape approximations are made in
the LMTO-FP. The atomic relaxations around the $F$ center
cannot be calculated by means of the ASA. We therefore performed
the structural optimization by
minimizing the LMTO-FP total energy calculated for a supercell. A similar
method was used in earlier LMTO-FP simulations of defects in KCl
(Ref. \onlinecite{chr1}) and MgO.\cite{chr2}
The supercell used in the present work contains 40 atoms
for the perfect KNbO$_3$ (8 formula units) and 39 atoms plus one empty
sphere in the $F$ center case.

\subsection{INDO}
The INDO calculation scheme and the computer code
{\sc CKUSTERD} were discussed in
detail in Refs.\onlinecite{indo1,indo2} and \onlinecite{egl1}. With this code
it is possible to perform both
cluster and periodic system calculations containing hundreds of atoms as
well as to carry out automated geometry optimization which is especially
important in defect calculations. In the periodic calculations the so-called
large unit cell (LUC) model is used. \cite{evar} Its idea is to perform the
electronic structure calculations for an extended unit cell at the wave
vector {\bf k}=0 in the narrowed Brillouin zone (BZ) which is equivalent to
band calculations at several special points of the normal BZ, transforming
to the narrow BZ center after the corresponding extension of the primitive
unit cell. In the KNbO$_3$ case the unit cell contains five atoms whereas
the $2\!\times\!2\!\times\!2$ extended (super)cell consists of 40 atoms.
A detailed analysis of the KNbO$_3$ parametrization for the INDO method is
presented in Ref.\onlinecite{egl1}. In that work
considerable covalency was found of the chemical bonding in pure KNbO$_3$.
The effective charges found from Mulliken population analysis
are (in units of $|e|$) $+0.543$ for K, $+2.019$ for Nb, and $-0.854$ for O,
which is very different from the expectation of the generally accepted
$ionic$ model: +1, +5, and $-$2, respectively. This is in agreement with the
effective atomic charges found in an experimental study of LiNbO$_3$.\cite
{lines} Our results emphasize a high degree of covalency of the Nb-O bond as
may be expected from intuitive electronegativity considerations and the fact
of a strong overlap between O $2p$ and Nb $4d$ orbitals and partial
densities of states. We discuss below how  covalency of the
chemical bonding may   have  important consequences for the physics of
defects in ferroelectrics, in particular for the $F$ centers.

To simulate $F$ centers, we started with a 40-atom supercell with one of the O
atoms removed. In the cubic phase all O atoms are equivalent and have
local symmetry C$_{4v}$ whereas in the orthorhombic phase there are $two$
kinds of nonequivalent O atoms whose symmetry is lower, $C_{2v}$ or $C_s$
(see below). After the O atom is removed, the atomic configuration of
surrounding atoms is reoptimized via a search of the total energy minimum as
a function of the atomic displacements from regular lattice sites.
Calculation of the adiabatic energy curves for the ground and excited states
permits us to find the optical absorption energy using the so-called
$\Delta$ self-consistent-field (SCF) procedure according to which the
$E_{abs}$ sought for is the difference
of the total energies for the ground and excited state with the defect
geometry of the ground state unchanged (vertical optical transition).

To extend the basis set in the $F$ center calculation, additional 1$s$, 2$p$
atomic orbitals were centered on the O vacancy. Their parameters were chosen
close to those used in the $F$ center calculations in MgO crystal:\cite{rob}
the orbital exponent $\zeta$($1s$)=0.65 (a.u.$^{-1}$),
$\zeta $($2p$)=0.50 (a.u$^{-1}$), the relevant electronegativities are
zero ($1s$) and $-$3 eV ($2p$), and the bonding parameters $\beta$=0
for both $1s$ and $2p$ orbitals, respectively.  During the defect geometry
optimization, we make no {\it \ a priori} assumptions on the electron
density distribution.

\section{ground-state properties}
\label{sec:res1}
\subsection{LDA calculations}
\begin{figure}
\epsfxsize=8.0cm
\centerline{\epsfbox{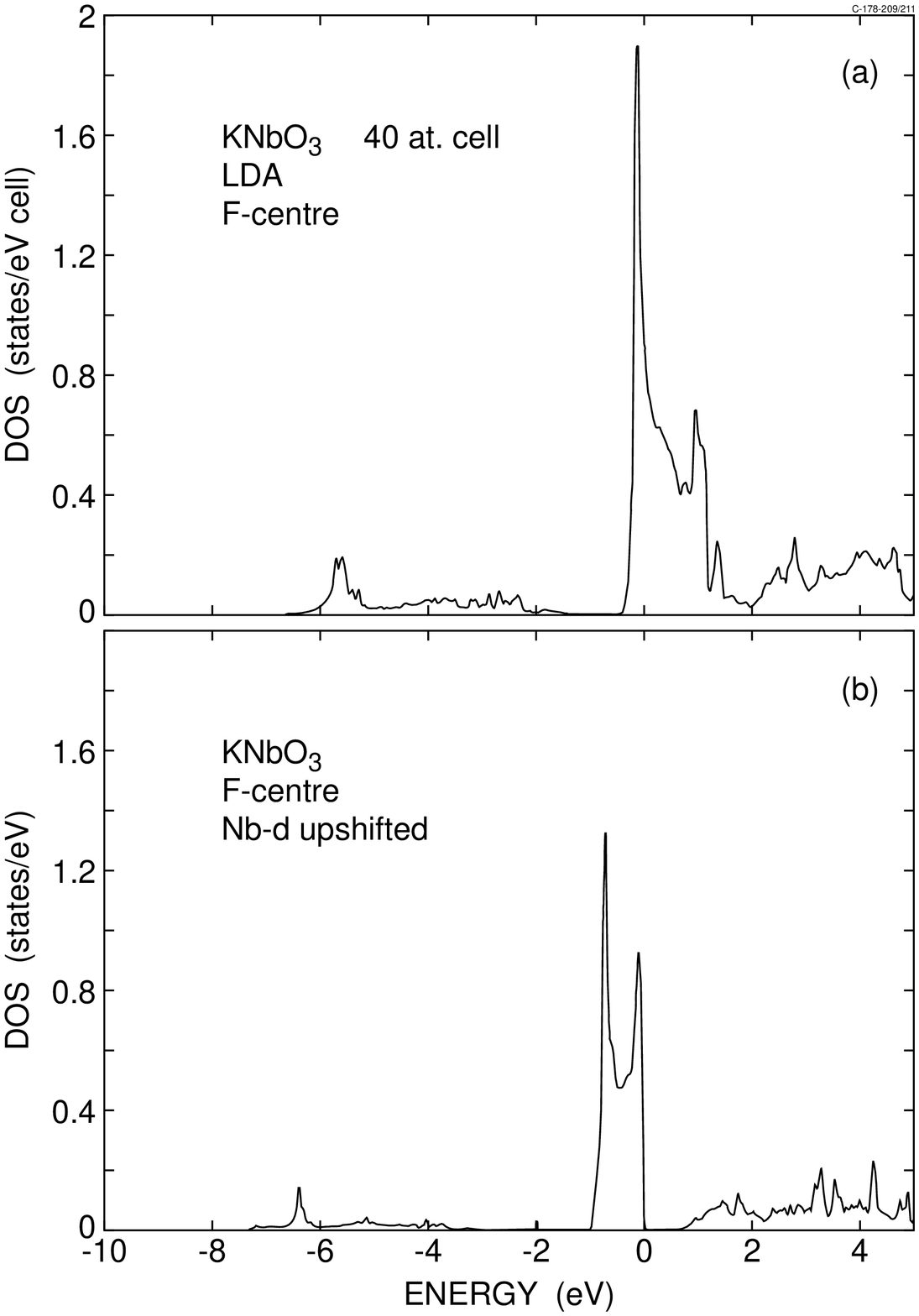}}
\vspace*{0.3cm}
\caption{(a) Partial $F$-center density of states for KNbO$_3$ calculated
within the LDA using a 39-atom supercell. Energies are given relative
to  the Fermi level. This band structure is metallic because the
finite size of the supercell causes the defect band  to have a width of
$\approx$0.8 eV so that it overlaps with the too-low-lying conduction
band (``LDA gap error'').
(b) As in (a) but with application of an external potential that upshifts
the Nb $d$ states in energy by $\approx$2.5 eV. Zero energy corresponds to
the top of the $F$ center band.}
\label{fniels1}
\end{figure}

\begin{figure}
\epsfxsize=8.0cm
\centerline{\epsfbox{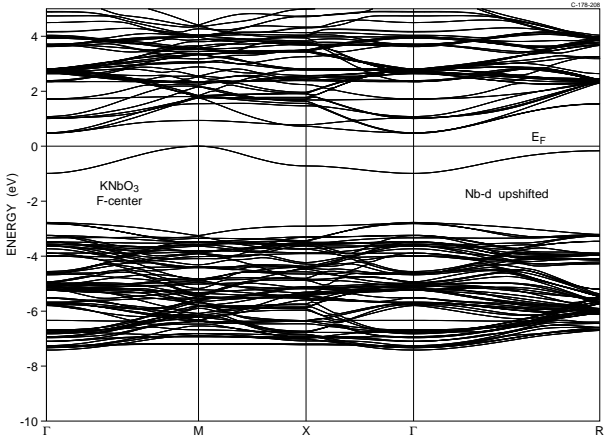}}
\vspace*{0.3cm}
\caption{Calculated band structure for the 39-atom supercell
modeling the $F$ center. The Nb $d$ states are shifted up as in Fig. 1(b).
The horizontal line (``$E_F$'') indicates the top of the defect band,
the highest occupied state.}
\label{fniels2}
\end{figure}

The band structure as derived from the straight LDA underestimates the gap
between occupied and empty states. Since the supercell which we use is
rather small (40 atomic sites), the defect states form a band of a finite
width ($\approx$0.8 eV). These two effects cause the defect band to
overlap with the conduction band, and the supercell calculations within the
LDA predict KNbO$_3$ with the $F$ centers to be a metal. This is
seen from Fig. \ref{fniels1}(a). This affects the charge distribution.
The number of electrons in the vacancy '`atomic sphere'' is  0.24, much
smaller than obtained from the INDO (0.6; see below). We can
artificially increase the gap by applying an upshift in each iteration to the
Nb $d$ states. Figure \ref{fniels1}(b) shows such a calculation, and
now the defect band lies completely within the gap. The energy scale
was chosen so that the top of the defect band (the highest occupied state)
is at $E$=0. The band plot, Fig. \ref{fniels2} also shows that the defect
band disperses over $\approx$0.8 eV due to the small supercell size.
After adjustment of the gap the (self-consistent) calculation yields
0.6 electrons inside the O vacancy sphere,
i.e., more than twice the amount found before and very close to the INDO
calculation for the {\it relaxed} structure. When relaxations (see below) are
included, the LMTO calculation yields a lower electron number in the vacancy
sphere. This is simply caused by the outward motion of the nearest neighbors
(Nb).

\begin{figure}
\epsfxsize=8.0cm
\centerline{\epsfbox{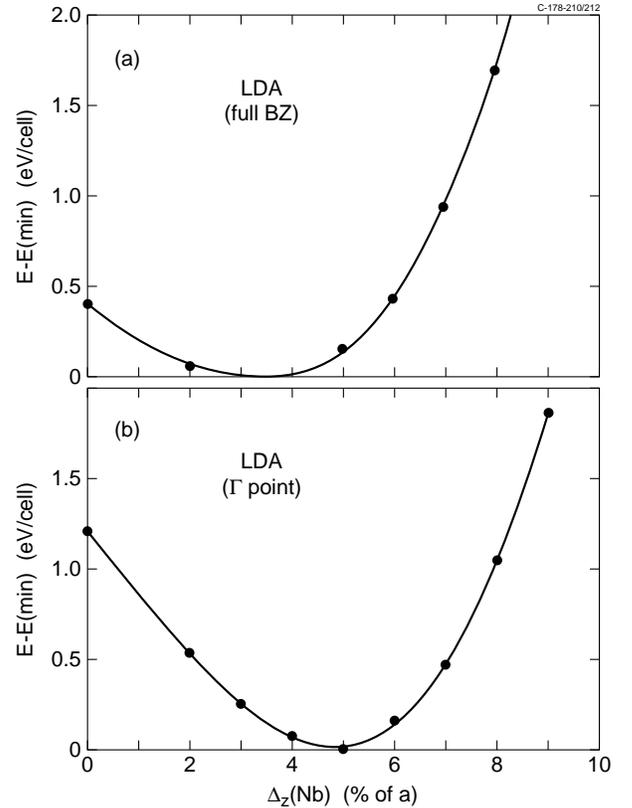}}
\vspace*{0.5cm}
\caption{(a). LMTO-FP calculation within the LDA of the total energy vs
the outward displacement, $\delta_z$, of the two Nb atoms closest to the
O vacancy. Note that no corrections were made in this case to ensure that
 the defect band lies entirely inside the gap. Integration in {\bf k}
space used 40 points in the irreducible part of the BZ.
(b). Same as (a) but only a single $\Gamma$ point was here included in the
{\bf k}-space sampling, in order to simulate the LUC used in the INDO
calculations. The curve is a high-order polynomial fit.}
\label{fniels3}
\end{figure}

\begin{table*}
\caption{Coordinates of three kinds of 14 atoms surrounding $F$ center
placed at the coordinate origin
in the cubic phase (in units of the lattice constant $a_0$=4.016 \AA ).}
\begin{tabular}{ccccc}
Atom &\multicolumn{3}{c}{Lattice coordinates} & Displacement \\
\hline
\multicolumn{5}{c}{without additional AO}\\
Nb &0    &0  &$\pm$($\frac{1}{2}+\Delta_z$)
 &$\Delta_z=0.08320$ \\
K  &$\pm$($\frac{1}{2}+\Delta_x$) &$\pm$($\frac{1}{2}+\Delta_y$)   &0
   &$\Delta_x=0.01324$, $\Delta_y=0.01324$  \\
O  &$\pm$($\frac{1}{2}-\Delta_x$) &0 &$\pm$($\frac{1}{2}-\Delta_z$)
   &$\Delta_x=0.02372$, $\Delta_z=0.00447$ \\
O  &0 &$\pm$($\frac{1}{2}-\Delta_y$) &$\pm$($\frac{1}{2}-\Delta_z$)
   &$\Delta_y=0.02372$, $\Delta_z=0.00447$  \\
\hline
\multicolumn{5}{c}{with additional AO}\\
Nb &0 &0 &$\pm$($\frac{1}{2}+\Delta_z$)
   &$\Delta_z=0.06550$ \\
K  &$\pm$($\frac{1}{2}+\Delta_x$) &$\pm$($\frac{1}{2}+\Delta_y$)    &0
   &$\Delta_x=0.00920$, $\Delta_y=0.00920$  \\
O  &$\pm$($\frac{1}{2}-\Delta_x$) &0 &$\pm$($\frac{1}{2}-\Delta_z$)
   &$\Delta_x=0.01920$,  $\Delta_z=0.00334$ \\
O  &0 &$\pm$($\frac{1}{2}-\Delta_y$) &$\pm$($\frac{1}{2}-\Delta_z$)
   &$\Delta_y=0.01920$,  $\Delta_z=0.00334$ \\
\end{tabular}
\label{tabek1}
\end{table*}

The relaxation of atoms surrounding the $F$ center was first calculated within
the LDA without any attempt to correct for the effect of the overlap between
the defect state band and the conduction band. First, the nearest-neighbor
Nb atoms were relaxed, and the result is illustrated in Fig. \ref
{fniels3}(a) which shows the total energy as a function of the outward
displacement, $\Delta_z$, of the Nb atom from its equilibrium position in the
bulk crystal. The relaxed atomic positions correspond to $\Delta _z$=3.5\%
of the lattice constant of $a_0$=4.016 \AA . This is about half the relaxation
found in the INDO calculation (see below). Further, the relaxation energy
found here, $\approx $ 0.5 eV, is much smaller than the value of 3.7 eV
obtained in the INDO calculation.

We do not wish to rely on total energy calculations where we applied 2.5 eV
upshift to the Nb $d$ bands. However, self-consistent calculations using
supercells large enough for obtaining a small width of the defect band
nonoverlapping with the LDA conduction band are impractical. An approximate
calculation was instead made of using the same supercell size as before but
sampling only the $\Gamma $ point of the BZ in the {\bf k}-space
integration. This is the point where the defect band has its minimum energy
and lies inside the gap, even in the LDA calculation. Further, this sampling
is the same as used in the INDO calculation described below [the so-called
large unit cell (LUC) model]. As expected, this changes the charge distribution
and the atomic relaxations. The  value of $\Delta _z$= 4.8\%
[Fig. \ref{fniels3})(b)] is closer to that obtained in the INDO calculation.
The relaxation energy  1.2 eV is also somewhat closer to the INDO result.

\begin{table*}[t]
\caption{Atomic coordinates for a perfect orthorhombic phase of KNbO$_3$
[in terms of lattice parameters $a$=3.973 \AA\ , $b$=5.695 \AA\ ,
$c$=5.721 \AA\
(INDO calculated in Ref. \protect\onlinecite{egl1})],
as well as positions of the two
Nb atoms nearest to the $F$ center related to a O(I) vacancy (point
symmetry $C_{2v}$) and  O(II) vacancy (symmetry $C_s$,
respectively). The sketch of the orthorhombic unit cell is shown in
Fig. \protect\ref{fek1}.}
\begin{tabular}{lr@{$+$}lr@{$+$}lr@{$+$}ldd}
Atom & \multicolumn{2}{c}{$a$} & \multicolumn{2}{c}{$b$} &
\multicolumn{2}{c}{$c$} & \multicolumn{2}{c}{$\Delta$} \\
\hline
\multicolumn{9}{c}{Perfect crystal} \\
K     & \multicolumn{2}{c}{0} & \multicolumn{2}{c}{0} &
\multicolumn{2}{c}{$\Delta_z$} & \multicolumn{2}{c}{$\Delta_z=0.0209$} \\
Nb    & \multicolumn{2}{c}{$\frac{1}{2}$} & \multicolumn{2}{c}{0} &
\multicolumn{2}{c}{$\frac{1}{2}$} \\
O$_{\text{I}}$  & \multicolumn{2}{c}{0} &
\multicolumn{2}{c}{0} & \multicolumn{2}{c}{$\frac{1}{2}+\Delta_z$} &
\multicolumn{2}{c}{$\Delta_z=0.0347$} \\
O$_{\text{II}}$  & \multicolumn{2}{c}{$\frac{1}{2}$} &
\multicolumn{2}{c}{$\frac{1}{4}+\Delta_y$} &
\multicolumn{2}{c}{$\frac{1}{4}+\Delta_z$} &
   $\Delta_y$=$-$0.0028 & $\Delta_z$=0.0347 \\
O$_{\text{II}}$  & \multicolumn{2}{c}{$\frac{1}{2}$} &
\multicolumn{2}{c}{$\frac{3}{4}-\Delta_y$} &
\multicolumn{2}{c}{$\frac{1}{4}+\Delta_z$} &
   $\Delta_y$=$-$0.0028 & $\Delta_z$=0.0347 \\
\multicolumn{9}{c}{$F$ centers of $C_{2v}$ symmetry: no additional AO} \\
Nb(I)    &  $\frac{1}{2}$&$\Delta_x$ & \multicolumn{2}{c}{0}
   &  $\frac{1}{2}$&$\Delta_z$ &
$\Delta_x$=0.08333 & $\Delta_z$=0.0132 \\
Nb(II)   & $-\frac{1}{2}$&$\Delta_x$ & \multicolumn{2}{c}{0}
   &  $\frac{1}{2}$&$\Delta_z$ &
   $\Delta_x$=$-$0.08333 & $\Delta_z$=0.0132 \\
\multicolumn{9}{c}{$F$ centers of $C_{2v}$ symmetry: with additional AO} \\
Nb(I)    &  $\frac{1}{2}$&$\Delta_x$ & \multicolumn{2}{c}{0}
  &  $\frac{1}{2}$&$\Delta_z$ &
   $\Delta_x$=0.0656 & $\Delta_z$=0.0106 \\
Nb(II)   & $-\frac{1}{2}$&$\Delta_x$ & \multicolumn{2}{c}{0}
  &  $\frac{1}{2}$&$\Delta_z$ &
   $\Delta_x$=$-$0.0656 & $\Delta_z$=0.0106 \\
\multicolumn{9}{c}{$F$ centers of $C_s$ symmetry: no additional AO} \\
Nb(I)    &  \multicolumn{2}{c}{~$\frac{1}{2}$} &
\multicolumn{2}{c}{$\Delta_y$} &
  $\frac{1}{2}$&$\Delta_z$ &
  $\Delta_y$=$-$0.0468 & $\Delta_z$=0.02502 \\
Nb(II)   & \multicolumn{2}{c}{$-\frac{1}{2}$} &
  $\frac{1}{2}$&$\Delta_y$ &
\multicolumn{2}{c}{$\Delta_z$} &
  $\Delta_y$=0.0244 & $\Delta_z$=$-$0.03735\\
\multicolumn{9}{c}{$F$ centers of $C_s$ symmetry: with additional AO} \\
Nb(I)    &  \multicolumn{2}{c}{~$\frac{1}{2}$} &
\multicolumn{2}{c}{$\Delta_y$} &
  $\frac{1}{2}$&$\Delta_z$ &
  $\Delta_y$=$-$0.0361 & $\Delta_z$=0.0192 \\
Nb(II)   & \multicolumn{2}{c}{$-\frac{1}{2}$} &
  $\frac{1}{2}$&$\Delta_y$ &
\multicolumn{2}{c}{$\Delta_z$} &
  $\Delta_y$=0.0187 & $\Delta_z$=$-$0.0286\\
\end{tabular}
\label{tab:ek2}
\end{table*}

\subsection{INDO}
\subsubsection{Cubic phase}
The positions of 14 atoms surrounding the $F$ center in a cubic phase after
lattice relaxation to the miminum of the total energy are given in Table \ref
{tabek1} for the two cases of with an additional atomic orbital (AO)
on vacancy site, and without the AO.
The conclusion is that the largest relaxation is exhibited by the two
nearest Nb atoms which are strongly (by 6.5\% of a$_0$) displaced
$outwards$ the O vacancy along the $z$ axis. This
is accompanied by a much smaller, 0.9\% outward displacement of K atoms and
by a 1.9\% inward displacement of O atoms. The two Nb atoms give the largest
\mbox{($\approx$ 80\%)} contribution to the lattice relaxation energy (3.7 eV)
whereas O atoms give the most of the rest energy gain of 1 eV.

The analysis of the effective charges of atoms surrounding the $F$ center
shows that of the two electrons associated with the removed O atom only
$\approx -0.6 |e|$ is localized inside $V_{\mbox{\tiny O}}$ and
a similar amount of the electron density is localized on the two nearest
Nb atoms. The $F$ center produces a local energy level, which lies
$\approx$ 0.6 eV above the top of the valence band. Its molecular orbital
contains primarily contribution from the atomic orbitals of the two nearest
Nb atoms.

\subsubsection{Orthorhombic phase}
\begin{figure}
\epsfxsize=8.0cm
\centerline{\epsfbox{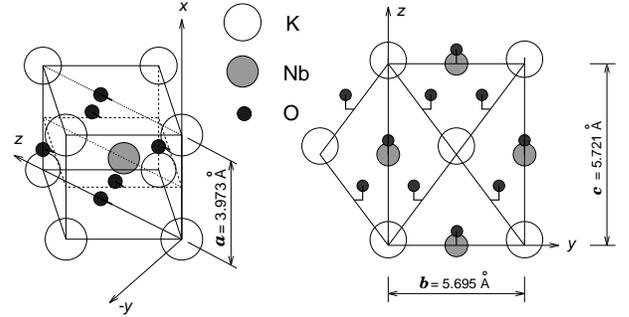}}
\vspace*{0.5cm}
\caption{
Orthorhombic phase of KNbO$_3$: equilibrium geometry. Atomic
coordinates are given in Table \protect\ref{tab:ek2}.}
\label{fek1}
\end{figure}

The orthorhombic phase of KNbO$_3$ is important since it is stable in a
broad temperature range around room temperature and thus is subject to most
studies and practical applications. The atomic positions for the $perfect$
orthorhombic cell shown in Fig.~\ref{fek1} were calculated earlier
\cite{egl1} and included in Table~\ref{tab:ek2}. Their  agreement
with the experimental data \cite{hewat} is very good.\cite{egl1}

The displacements of Nb atoms nearest to $V_{\mbox{\tiny O}}$ are given
in the same Table \ref{tab:ek2} for the both kinds of $F$ centers existing
in this phase and, again, with and without atomic orbitals centered
at the O vacancy. In fact, Nb displacements are very similar in magnitude
(6.6\%) and also close to those found for the cubic phase.
The relevant relaxation energies are considerable
(3.6 eV) and nearly the same as that for Nb relaxation found in the cubic
phase.

\section{Optical properties}
\label{sec:opt}
Because of the $C_{4v}$ local symmetry of the $F$ centers in the KNbO$_3$
cubic phase, its excited state splits into two levels, one of which remains
twofold degenerate. Our $\Delta$SCF calculations predict the two relevant
absorption bands: at 2.73 eV and 2.97 eV (Table \ref{tab:ek3}). Neglect of
the additional orbitals centered on the $V_{\mbox{\tiny O}}$ slightly
affect this result, the relevant absorption energies turn out to be
2.67 eV and 3.02 eV, respectively.

\begin{table}[t]
\caption{Calculated absorption ($E_{\mbox{\tiny abs}}$, eV)
and Nb-atom relaxation ($E_{\mbox{\tiny rel}}$, eV)
energies for the $F$ center for the cubic and the orthorhombic phases}
\begin{tabular}{lllll}
 Symmetry, phase & &$E_{\mbox{\tiny abs}}$  &  &$E_{\mbox{\tiny rel}}$ \\
\hline
 $C_{4v}$, cubic        &2.73  &2.97  &--    & 3.7\\
 $C_s$, orthorhombic    & 2.56 & 3.03 & 3.10 & 3.6\\
 $C_{2v}$, orthorhombic & 2.72 & 3.04 & 3.11 & 3.6\\
\end{tabular}
\label{tab:ek3}
\end{table}

Around room temperature in the orthorhombic phase there exist two kinds
of $F$ centers associated with two non-equivalent O atoms revealing
the $C_{2v}$ and $C_s$ symmetry; cf. Table \ref{tab:ek2}. The corresponding
three absorption bands for each of them are shown in Table \ref{tab:ek3}.
Their difference is the largest for the lowest-energy bands (0.16 eV) and
very small for other two bands.

\section{Summary}
\label{sec:conc}
Our results are in a sharp contrast with what is known for $F$ centers in
ionic crystals (in particular, in MgO and alkali halides \cite{kot1,craw})
where the two electrons are well localized by the $V_{\mbox{\tiny O}}$ in the
$F$ center  ground state.
Evidently, this discrepancy arises from a considerable degree of {\it %
covalency of the chemical bonding in KNbO$_3$} which is neglected in all
previous models of defects in this material (as well as similar ABO$_3$
perovskites, e.g. Refs. \onlinecite{swe,varn}); the only exception
known to us is an $X_\alpha$ cluster calculation of $F$ centers
in LiNbO$_3$. \cite{leo}

Electron defects similar to what we have observed here are known,
in particular, in partly covalent SiO$_2$ crystals (e.g., in the so-called
$E^{\prime}_1$ center an electron is also not localized inside
$V_{\mbox{\tiny O}}$
but its wave function mainly overlaps with the $sp^3$
orbital centered on the neighboring Si atom \cite{yip}).

We found that the ground state of the $F$ center is associated with a strong
$symmetrical$ relaxation of the two nearest Nb atoms outwards relative
to the O vacancy.
These Nb atoms remain to be identical; i.e., we did not see formation of
dipole moments of the Nb$_1$-$V_{\mbox{\tiny O}}$-Nb$_2$ type,
as suggested in Ref. \onlinecite{pros}.
Note that the relevant relaxation energy is several eV which is
typical for many point defects in ionic and partly ionic solids. Its
magnitude is by several orders of magnitude larger than the tiny energy gain
due to the phase transitions (meV per cell).

We presented a strong argument that the 2.7 eV absorption band observed in
electron irradiated crystals \cite{hod} could be due to the $F$ centers, and
predicted existence of two additional absorption bands
(at 3.04 eV and 3.11 eV) for the same defect in the orthorhombic phase
of KNbO$_3$ (see also discussion in Ref. \onlinecite{F+}).
At higher temperatures where the cubic phase is stable,
the latter two energies which hardly could be separated experimentally
because of the large half-width of absorption bands, degenerate into a
single, double-degenerate level at 2.97 eV.

Our results also suggest that the blue-light-induced-IR
absorption effect \cite{polzik}
mentioned in the Introduction  could be triggered by
the $F$ center absorption which  may lead to
its subsequent ionization where an electron is transferred
to the conduction band. The retrapping of this electron by
another defect is then responsible for the infrared absorption
when the retrapped electron is excited to the conduction band.
We recall that the UV excitation (blue-light) energy used in these experiments
is very close to our calculated absorption energy of the $F$ centers.

The reason for the discrepancy in LDA and INDO relaxation energies
for the $F$ center needs further study; probably it arises due to strong
electron correlation effects which are crudely approximated in the
two methods in quite different ways.

\acknowledgements
R.E. has been supported by the Nieders\"achsisches Ministerium f\"ur
Wissenschaft und Kultur. A.P. has been supported by the Deutsche
Forschungsgemeinschaft (SFB~225). E.K. greatly appreciates the financial
support from Danish Natural Science Research Council (Contract No. 9600998)
and Latvian National Program on ``New Materials for Microelectronics''.
The authors are grateful to A.I. Popov and E. Stefanovich for fruitful
discussions.

\end{document}